\let\orilabel\label
\documentclass[aps,pra,preprint,a4paper,showpacs,showkeys,superscriptaddress,nofootinbib]{revtex4-1}
\let\label\orilabel

\usepackage{latexsym}
\usepackage{amsmath,amssymb,mathrsfs}
\usepackage{mathtools}
\usepackage{graphicx}
\usepackage{subfigure}
\usepackage{xcolor}
\usepackage{physics}
\usepackage{cancel}
\usepackage{tensor}
\usepackage{tikz}
\usepackage{multirow,tabularx}

\newcolumntype{C}[1]{>{\centering\arraybackslash}p{#1}}
\usepackage{hyperref}     
\hypersetup{colorlinks,%
citecolor=blue,%
linkcolor=cyan,%
}
\usepackage[titletoc]{appendix}
\usepackage{enumerate}
\graphicspath{{./Figures/}}
\usetikzlibrary{decorations.pathmorphing,decorations.pathreplacing,decorations.shapes,calc}
\begin{document}


\title{Einstein ring of dust shells with quantum hair}

\author{Sojeong Cheong}%
\email[]{jsquare@sogang.ac.kr}%
\affiliation{Department of Physics, Sogang University, Seoul, 04107, Republic of Korea}%
\affiliation{Center for Quantum Spacetime, Sogang University, Seoul, 04107, Republic of Korea}%

\author{Wontae Kim}%
\email[]{wtkim@sogang.ac.kr}%
\affiliation{Department of Physics, Sogang University, Seoul, 04107, Republic of Korea}%
\affiliation{Center for Quantum Spacetime, Sogang University, Seoul, 04107, Republic of Korea}%

\author{Mungon Nam}%
\email[]{clrchr0909@sogang.ac.kr}%
\affiliation{Research Institute for Basic Science, Sogang University, Seoul, 04107, Republic of Korea}%

\date{\today}

\begin{abstract}
	The information about the internal structure of a compact object is classically inaccessible to external observers.
    In this paper, we investigate how quantum corrections to gravitational fields can reveal the internal structure of compact objects composed of dust shells.
	Using an effective field theory approach to incorporate quantum corrections up to second order in curvature, we derive a quantum‐corrected metric for $N$ uniformly spaced shells with equal surface mass density and then examine
how these corrections manifest in the deflection angle for gravitational lensing.
	In particular, we mainly investigate quantum-corrected astrophysical observables such as the Einstein ring and image magnification.
	Compared to the classical scenario, the deflection angle and the corresponding Einstein angle differ by a term that depends explicitly on the number of dust shells, which play the role of quantum hair.
	Specifically, the quantum correction to them diminishes as $N$ increases, yet a finite deviation from the classical result remains even in the
    continuum limit $N\to\infty$.
	Consequently, our results show that the internal structures of compact objects with identical mass and radius can be distinguished by quantum hair through their lensing observables.
\end{abstract}


\maketitle
\raggedbottom

\section{introduction}
\label{sec:introduction}

The no-hair theorem states that a stationary black hole is completely characterized by three parameters: its mass, charge, and angular momentum~\cite{Israel:1967wq,Israel:1967za,Carter:1971zc,Robinson:1975bv}.
In other words, any additional properties of the collapsing matter are hidden behind the event horizon and cannot be accessed by external observers.
It is important to note, however, that the no-hair theorem relies on classical solutions to the Einstein field equations.
The theorem does not exclude the possibility that stationary black holes may possess additional quantum characteristics beyond the classical framework, which are deeply connected to the concept of ``quantum hair.''

In connection with quantum effects in gravitational systems, one approach is to use the unique effective action method, developed by Barvinsky and Vilkovisky~\cite{Barvinsky:1983vpp,Barvinsky:1985an,Barvinsky:1987uw,Barvinsky:1990up,Barvinsky:1990uq}, which expresses the one-loop effective action in terms of curvatures and covariant derivatives.
They extended the Schwinger-DeWitt technique~\cite{Schwinger:1951nm,DeWitt:1964mxt} into a covariant perturbation theory by expanding the effective action in powers of curvature while treating the zeroth-order term as flat spacetime.
This covariant perturbation approach provides a powerful tool for examining how quantum effects manifest in gravitational systems.
In this context, this method has been employed in a wide variety of studies, such as those on gravitational waves
~\cite{Calmet:2016sba,Calmet:2017rxl,Calmet:2018rkj}, black holes~\cite{Calmet:2017qqa,Calmet:2018elv,Calmet:2020tlj,Calmet:2021lny,Xiao:2021zly,Delgado:2022pcc},
and
stars~\cite{Calmet:2019eof,Calmet:2021stu,Calmet:2022bpo,Calmet:2023met,Cheong:2023oik,Perrucci:2024qrr}.

There is an intriguing distinction between Refs.~\cite{Calmet:2017qqa,Calmet:2018elv,Calmet:2020tlj,Calmet:2021lny,Xiao:2021zly,Delgado:2022pcc} and Refs.~\cite{Calmet:2019eof,Calmet:2021stu,Calmet:2022bpo,Calmet:2023met,Cheong:2023oik,Perrucci:2024qrr} in that, at quadratic order in curvature,
quantum corrections to the metric emerge not for black holes but for stars.
The quantum corrections to the metric come from local and nonlocal operators associated with the Ricci scalar and Ricci tensor in the effective action.
In particular, for the Schwarzschild black hole, there are no quantum corrections to the Schwarzschild metric~\cite{Calmet:2017qqa,Calmet:2018elv,Calmet:2020tlj,Calmet:2021lny,Xiao:2021zly,Delgado:2022pcc}.
For a star, however, the nonlocal operators in the effective action non-trivially act on the Ricci scalar and Ricci tensor
because of the mass distribution of the star. So, the exterior metric receives quantum effects depending on its internal structure~\cite{Calmet:2019eof,Calmet:2021stu,Calmet:2022bpo,Calmet:2023met,Cheong:2023oik,Perrucci:2024qrr}.
This implies that even for stars with identical mass and radius, an external observer could distinguish between them through quantum effects, indicating the breakdown of Birkhoff's theorem.
Quantum corrections to the metric of a star have attracted significant interest in the study of its internal structure with quantum hair~\cite{Calmet:2021stu,Calmet:2022bpo,Calmet:2023met}.

In particular, in a multi-layered star model~\cite{Calmet:2021stu}, it was first demonstrated that the internal structure of a star can play the role of quantum hair.
It was shown that, in a configuration consisting of two concentric solid dust spheres with distinct constant mass densities, the dependence on the mass and radius of them appears in the $r^{-5}$ terms of the quantum-corrected metric.
This result clearly differs from that obtained from a single solid sphere model.
Accordingly, the internal structure of the multi-layered solid spheres was identified as quantum hair, enabling an external observer to distinguish between different stars.
Subsequent studies have further explored quantum effects in the metric arising from various internal structures~\cite{Cheong:2023oik,Perrucci:2024qrr}.

On the other hand, the gravastar was proposed as an alternative to a black hole, consisting of three regions characterized by the equations of state of its internal matter: an inner region with negative pressure, an intermediate shell region, and an outer region described by the Schwarzschild metric~\cite{Mazur:2001fv,Mottola2023}.
When quantum effects are taken into account, the state parameter $w$ turned out to be quantum hair, which modified the photon sphere radius and the deflection angle~\cite{Perrucci:2024qrr}.
If we take the limit of $w=0$ where quantum hair disappears,
the star becomes a solid dust sphere.
In this case, for a more complicated internal structure,
one can consider a discretized star as a series of
$N$ dust shells, each with identical surface mass density, while recovering the solid dust sphere model in the limit $N\to\infty$.
It can be expected that the number of dust shells will play the role of quantum hair.
Thus, it would be interesting to study how the internal structure can be read off through astrophysical observables related to quantum hair.

In this paper, based on the Barvinsky-Vilkovisky method~\cite{Barvinsky:1983vpp,Barvinsky:1985an,Barvinsky:1987uw,Barvinsky:1990up,Barvinsky:1990uq}, we
consider the quantum effective action that includes the classical matter contribution for the $N$ dust shells, expanded up to second order in curvature.
Solving the equations of motion perturbatively in powers of the gravitational constant, we derive the quantum-corrected metric around flat spacetime.
Note that according to the Barvinsky-Vilkovisky formalism, one should take the background as the flat metric although the Schwarzschild metric has been adopted as the background in the previous works~\cite{Calmet:2017qqa,Calmet:2019eof,Calmet:2020tlj,Xiao:2021zly,Cheong:2023oik,Perrucci:2024qrr,Calmet:2023met}.
Using the resultant metric, we investigate the internal structure of dust shells through the gravitational lensing effect~\cite{Liebes:1964zz,Schneider:1992bmb,Keeton:2005jd,Kundu:2024wyx,Vachher:2024ait}, with a particular focus on astrophysical observables such as the Einstein ring and the image magnification.
Compared to the classical scenario, the quantum-corrected Einstein ring and image magnification as well as the deflection angle
will be shown to depend on the internal structure of the compact object.

The organization of this paper is as follows.
In Sec.~\ref{sec:corrections}, we derive the quantum-corrected metric for the uniformly spaced $N$ dust shells.
In Sec.~\ref{sec:deflection_angles}, we investigate the gravitational lensing effect by calculating the deflection angle in the quantum-corrected metric, showing that it differs from the classical result.
In Sec.~\ref{sec:numerics}, we examine the lensing observables by numerically analyzing the Einstein ring and the magnification of an image.
The conclusion and discussion are presented in Sec.~\ref{sec:conclusion}.

\section{Quantum-corrected metric of \texorpdfstring{$N$}{N} dust shells}
\label{sec:corrections}
We begin with the bare action up to second order in curvature as
\begin{equation}
	\label{}
	S_{\rm bare} = \int\dd[4]x \sqrt{-g} \left[ \frac{1}{16\pi G}\mathcal{R} + c_{10} \mathcal{R}^2 + c_{20}\mathcal{R}_{\mu\nu}\mathcal{R}^{\mu\nu} + c_{30}\mathcal{R}_{\mu\nu\kappa\lambda}\mathcal{R}^{\mu\nu\kappa\lambda} \right] + S_{\rm shell} + S_{\rm field},
\end{equation}
where $S_{\rm shell}$ is the classical matter action for the $N$-shell structure, $S_{\rm field}$ is the action for massless fields (scalars, fermions, and vectors), and $c_{10}$, $c_{20}$, and $c_{30}$ are dimensionless bare coefficients.
Integrating out the massless fields in $S_{\rm field}$ yields the unique effective action of quantum gravity~\cite{Barvinsky:1983vpp,Barvinsky:1985an,Barvinsky:1987uw,Barvinsky:1990up,Barvinsky:1990uq}.
Up to second order in curvature, the resulting effective action can be written as~\cite{Barvinsky:1990up}
\begin{equation}
	\label{eq:eff action}
	S_{\rm eff} = \frac{1}{16\pi G}\int\dd[4]x \sqrt{-g} \mathcal{R} + S_{\rm shell} + \Gamma_{\rm L} + \Gamma_{\rm NL}.
\end{equation}
Here, the local part of the effective action $\Gamma_{\rm L}$ is
    \begin{equation}
    	\label{eq:local_action}
        \Gamma_{\rm L} = \int \dd[4]x \sqrt{-g} \left[ c_1(\mu) \mathcal{R}^2 + c_2(\mu) \mathcal{R}_{\mu\nu}\mathcal{R}^{\mu\nu} + c_3(\mu) \mathcal{R}_{\mu\nu\kappa\lambda}\mathcal{R}^{\mu\nu\kappa\lambda} \right]
    \end{equation}
and the nonlocal part $\Gamma_{\rm NL}$ is
    \begin{equation}
    	\label{eq:nonlocal_action}
        \Gamma_{\rm NL} = -\int \dd[4]x \sqrt{-g} \left[ \alpha \mathcal{R}\ln\left( -\frac{\square}{\mu^2}\right)\mathcal{R} + \beta \mathcal{R}_{\mu\nu}\ln\left( -\frac{\square}{\mu^2} \right)\mathcal{R}^{\mu\nu} + \gamma \mathcal{R}_{\mu\nu\kappa\lambda}\ln\left(-\frac{\square}{\mu^2} \right)\mathcal{R}^{\mu\nu\kappa\lambda} \right],
    \end{equation}
    where $\square = g^{\mu\nu}\nabla_\mu \nabla_\nu$ and $\mu$ is the renormalization scale parameter.
The renormalized coefficients $c_1(\mu)$, $c_2(\mu)$, and $c_3(\mu)$ in Eq.~\eqref{eq:local_action} satisfy the following renormalization group equations~\cite{Donoghue:2014yha}:
\begin{equation}
	\label{}
	\mu\dv{\mu}c_1(\mu) = -2\alpha,\quad \mu\dv{\mu}c_2(\mu) = -2\beta, \quad \mu\dv{\mu}c_3(\mu) = -2\gamma,
\end{equation}
where $\alpha$, $\beta$, and $\gamma$ are the Wilson coefficients depending on contributions from the massless fields.
Specifically,
\begin{align}
	\alpha &= \frac{1}{11520\pi^2}\left( 5(6\xi-1)^2\mathcal{N}_{s} - 5\mathcal{N}_{f} - 50 \mathcal{N}_{v} \right),\label{eq:alpha}\\
	\beta &= \frac{1}{11520\pi^2}\left( -2\mathcal{N}_{s} +8\mathcal{N}_{f} + 176 \mathcal{N}_{v} \right),\label{eq:beta}\\
	\gamma &= \frac{1}{11520\pi^2}\left( 2\mathcal{N}_{s} + 7\mathcal{N}_{f} - 26 \mathcal{N}_{v} \right),\label{eq:gamma}
\end{align}
where $\mathcal{N}_{s}$, $\mathcal{N}_{f}$, and $\mathcal{N}_{v}$ are the numbers of scalars, fermions, and vectors, respectively, and $\xi$ denotes a non-minimal scalar-curvature coupling~\cite{Birrell:1982ix,Buchbinder:1992rb}.
For simplicity, we assume a minimally coupled scalar field where $\mathcal{N}_{s} = 1$ , $\xi = 0$,  and $\mathcal{N}_{f} = \mathcal{N}_{v}  = 0$.

Varying the effective action \eqref{eq:eff action} with respect to the metric yields the following quantum-gravitational field equations:
    \begin{equation}
    \label{eq:eom}
      \mathcal{R}_{\mu\nu} - \frac{1}{2}g_{\mu\nu}\mathcal{R} + 16\pi G (H^{\rm L}_{\mu\nu}+H^{\rm NL}_{\mu\nu}) = 8\pi G T_{\mu\nu}^{\rm shell},
    \end{equation}
where $T_{\mu\nu}^{\rm shell} = -\frac{2}{\sqrt{-g}}\fdv{S_{\rm shell}}{g^{\mu\nu}}$.
The local contribution to the field equations $H^{\rm L}_{\mu\nu}$ is derived from $\Gamma_{\rm L}$ as
    \begin{align}\label{eq:local_eom}
      H^{\rm L}_{\mu\nu} &= \frac{1}{\sqrt{-g}}\fdv{\Gamma_{\rm L}}{g^{\mu\nu}} = 2\bar{c}_1 \left(\mathcal{R}\mathcal{R}_{\mu\nu} - \frac{1}{4}g_{\mu\nu}\mathcal{R}^2 + g_{\mu\nu}\square \mathcal{R} - \nabla_\mu\nabla_\nu \mathcal{R} \right) \nonumber\\
      &+ \bar{c}_2 \left( 2\mathcal{R}^\alpha_\mu\mathcal{R}_{\nu\alpha} - \frac{1}{2}g_{\mu\nu}\mathcal{R}_{\alpha\beta}\mathcal{R}^{\alpha\beta} + \square \mathcal{R}_{\mu\nu}  + \frac{1}{2}g_{\mu\nu}\square \mathcal{R} - \nabla_\alpha\nabla_\mu \mathcal{R}^\alpha_\nu - \nabla_\alpha\nabla_\nu \mathcal{R}^\alpha_\mu \right),
    \end{align}
and the nonlocal contribution $H^{\rm NL}_{\mu\nu}$ derived from $\Gamma_{\rm NL}$ is
    \begin{align}
    \label{eq:nonlocal_eom}
      H^{\rm NL}_{\mu\nu} &= \frac{1}{\sqrt{-g}}\fdv{\Gamma_{\rm NL}}{g^{\mu\nu}} = -2\bar{\alpha} \left(\mathcal{R}_{\mu\nu} - \frac{1}{4}g_{\mu\nu}\mathcal{R} + g_{\mu\nu}\square - \nabla_\mu\nabla_\nu \right)\ln\left(-\frac{\square}{\mu^2}\right)\mathcal{R}\nonumber\\
          &-\bar{\beta} \left(2\delta^{\alpha}_{(\mu}\mathcal{R}_{\nu)\beta} - \frac{1}{2}g_{\mu\nu}\mathcal{R}^{\alpha}_{\beta} +  \delta^\alpha_\mu g_{\nu\beta}\square + g_{\mu\nu}\nabla^\alpha\nabla_\beta - \delta^\alpha_\mu\nabla_\beta\nabla_\nu - \delta^\alpha_\nu\nabla_\beta\nabla_\mu \right)\ln\left(-\frac{\square}{\mu^2}\right)\mathcal{R}^\beta_\alpha .
    \end{align}
Note that the last terms in Eqs.~\eqref{eq:local_action} and \eqref{eq:nonlocal_action} are eliminated by the local and nonlocal Gauss–Bonnet theorem~\cite{Calmet:2018elv}, redefining the coefficients as $\bar{c}_1 = c_1 - c_3$, $\bar{c}_2 = c_2 + 4c_3$, $\bar{\alpha} = \alpha - \gamma$, and $\bar{\beta} = \beta + 4\gamma$.
As discussed in Refs.~\cite{Donoghue:2014yha,Donoghue:2015nba}, the variation of the logarithmic operators in Eq.~\eqref{eq:nonlocal_action} contributes only at orders higher than quadratic in curvature.
Thus, we neglect these higher-order contributions, which means that
$\ln(-\frac{\square}{\mu^2}) \approx \ln(-\frac{\bar{\square}}{\mu^2})$, where $\bar{\square} = \eta^{\mu\nu}\bar{\nabla}_{\mu} \bar{\nabla}_{\nu}$.

We now specify the energy-momentum tensor $T_{\mu\nu}^{\rm shell}$ for $N$ dust shells, arranged such that they have equal surface mass density $\sigma$ and are uniformly spaced in the radial direction.
For $n=1,2,\cdots,N$, the mass and radius of the $n$-th dust shell are then given by
\begin{equation}
	\label{}
	M_n = \frac{6M(N-n+1)^2}{N(N+1)(2N+1)},\quad  R_n = \frac{R(N-n+1)}{N},
\end{equation}
where the total mass $M= \Sigma^{N}_{n=1} M_n$ and $R_1>R_2>\cdots>R_N$.
For any $N$, the outermost radius is always $R_1$ which will be fixed as a constant value.
In this configuration, the energy-momentum tensor for the $N$ dust shells is expressed as
\begin{equation}
\label{}
T_{\mu\nu}^{\rm shell} =\rho(r)  U_{\mu}U_{\nu},
\end{equation}
where
\begin{equation}
\label{eq:density_rho}
\rho(r) = \sum^{N}_{n=1} \frac{M_n}{4\pi R_n^2}\delta(r-R_n) =\frac{3NM}{2\pi R^2 (N+1)(2N+1)} \sum^{N}_{n=1} \delta(r-R_n)
\end{equation}
with the constant surface mass density $\sigma=\frac{3NM}{2\pi R^2 (N+1)(2N+1)}$
and $U^{\mu}$ is the four-velocity of each dust shell, satisfying $U_{\mu}U^{\mu} = -1$.

Next, we perturb Eq.~\eqref{eq:eom} by expanding the metric in powers of the gravitational constant $G$ around flat spacetime as
\begin{equation}
	\label{eq:metric perturb}
	g_{\mu\nu} = \eta_{\mu\nu} + h_{\mu\nu}^{(1)} + h_{\mu\nu}^{(2)},
\end{equation}
where $h_{\mu\nu}^{(1)} = \mathcal{O}(G) $ and $h_{\mu\nu}^{(2)} = \mathcal{O}(G^2) $.
We choose a gauge condition such that $h_{\theta\theta}^{(1)} = h_{\theta\theta}^{(2)} = 0 $ and $ h_{\phi\phi}^{(1)} = h_{\phi\phi}^{(2)} = 0$, which ensures $g_{\theta\theta} = r^2$ and $g_{\phi\phi}=r^2\sin^2\theta$.
Under this gauge choice, Eq.~\eqref{eq:eom} simplifies to
\begin{equation}
	\label{eq:linear eom}
	\mathcal{G}_{\mu\nu}^{(1)}[h^{(1)}] = 8\pi G \rho(r) \delta^t_\mu \delta^t_\nu
\end{equation}
at the linear order in $G$, and
\begin{equation}
	\label{eq:quadratic eom}
	\mathcal{G}_{\mu\nu}^{(1)}[h^{(2)}] + \mathcal{G}_{\mu\nu}^{(2)}[h^{(1)}] + 16\pi G (H^{{\rm L}(1)}_{\mu\nu}[h^{(1)}]+H^{{\rm NL}(1)}_{\mu\nu}[h^{(1)}]) = -8\pi G \rho(r)h^{(1)}_{tt}\delta^t_\mu \delta^t_\nu
\end{equation}
at the quadratic order in $G$.
Here, $\mathcal{G}^{(1)}_{\mu\nu}$ and $\mathcal{G}^{(2)}_{\mu\nu}$ denote the linear and quadratic parts of the Einstein tensor in the perturbative expansion of the metric, respectively.
Explicitly, $\mathcal{G}^{(1)}_{\mu\nu}$ is given by
\begin{align}
	\mathcal{G}^{(1)}_{\mu\nu}[h] &= \mathcal{R}^{(1)}_{\mu\nu}[h] - \frac{1}{2}\eta_{\mu\nu}\mathcal{R}^{(1)}[h] \nonumber\\
	&= \bar{\nabla}^{\alpha}\bar{\nabla}_{(\mu}h_{\nu)\alpha}- \frac{1}{2}\bar{\nabla}_{\mu}\bar{\nabla}_{\nu}h_{\alpha}^{\alpha} -\frac{1}{2}\bar{\square}h_{\mu\nu} -\frac{1}{2}\eta_{\mu\nu}(\bar{\nabla}^{\alpha}\bar{\nabla}^{\beta} h_{\alpha\beta} - \bar{\square}h_{\alpha}^{\alpha})
\end{align}
and $\mathcal{G}^{(2)}_{\mu\nu}$ is
\begin{align}
	\mathcal{G}^{(2)}_{\mu\nu}[h] &= \frac{1}{2}h^{\alpha\beta}\bar{\nabla}_{\mu}\bar{\nabla}_{\nu}h_{\alpha\beta}+\frac{1}{4}\bar{\nabla}_{\mu}h_{\alpha\beta}\bar{\nabla}_{\nu}h^{\alpha\beta} + \bar{\nabla}^{\alpha}h^{\beta}_{\nu}\bar{\nabla}_{[\alpha}h_{\beta]\mu} - h^{\alpha\beta}\bar{\nabla}_{\alpha}\bar{\nabla}_{(\mu}h_{\nu)\beta} \nonumber\\
	&\quad + \frac{1}{2}\bar{\nabla}_{\alpha}(h^{\alpha\beta}\bar{\nabla}_{\beta}h_{\mu\nu})-\frac{1}{4}\bar{\nabla}_{\alpha}h_{\mu\nu}\bar{\nabla}^{\alpha}h^{\beta}_{\beta}-\left( \bar{\nabla}_{\beta}h^{\alpha\beta} - \frac{1}{2}\bar{\nabla}^{\alpha}h_{\beta}^{\beta} \right)\bar{\nabla}_{(\mu}h_{\nu)\alpha}\nonumber\\
	&\quad -\eta_{\mu\nu}\bigg( \frac{1}{2}h^{\alpha\beta}\bar{\square}h_{\alpha\beta} + \frac{1}{2}h^{\alpha\beta}\bar{\nabla}_{\alpha}\bar{\nabla}_{\beta}h^{\gamma}_{\gamma}-h^{\alpha\beta}\bar{\nabla}_{\alpha}\bar{\nabla}^{\gamma}h_{\beta\gamma} + \frac{3}{8}\bar{\nabla}_{\alpha}h_{\beta\gamma}\bar{\nabla}^{\alpha}h^{\beta\gamma} \nonumber\\
	&\quad - \frac{1}{4}\bar{\nabla}^{\alpha}h^{\beta\gamma}\bar{\nabla}_{\beta}h_{\alpha\gamma} + \frac{1}{2}\bar{\nabla}_{\alpha}h^{\alpha\beta}\bar{\nabla}_{\beta}h^{\gamma}_{\gamma}-\frac{1}{8}\bar{\nabla}_{\alpha}h_{\beta}^{\beta}\bar{\nabla}^{\alpha}h^{\gamma}_{\gamma}- \frac{1}{2}\bar{\nabla}_{\beta}h^{\alpha\beta}\bar{\nabla}^{\gamma}h_{\alpha\gamma} \bigg),
\end{align}
where $h^{\alpha\beta} = \eta^{\alpha\gamma}\eta^{\beta\delta}h_{\gamma\delta}$.
In addition, the linear components of $H^{{\rm L}}_{\mu\nu}$ and $H^{{\rm NL}}_{\mu\nu}$ in the perturbative expansion of the metric take the forms as
    \begin{align}\label{eq:local_eom_linear}
	H^{{\rm L}(1)}_{\mu\nu}[h] &= 2\bar{c}_1 \left(\eta_{\mu\nu}\bar{\square} \mathcal{R}^{(1)}[h] - \bar{\nabla}_\mu\bar{\nabla}_\nu \mathcal{R}^{(1)}[h] \right) \nonumber\\
	&\quad +\bar{c}_2 \left( \bar{\square} \mathcal{R}_{\mu\nu}^{(1)}[h]  + \frac{1}{2}\eta_{\mu\nu}\bar{\square} \mathcal{R}^{(1)}[h] - \bar{\nabla}^\alpha\bar{\nabla}_\mu \mathcal{R}^{(1)}_{\alpha\nu}[h] - \bar{\nabla}^\alpha\bar{\nabla}_\nu \mathcal{R}_{\alpha\mu}^{(1)}[h] \right),
\end{align}
and
\begin{align}
	\label{eq:nonlocal_eom_linear}
	H^{{\rm NL}(1)}_{\mu\nu}[h] &= -2\bar{\alpha} \left(\eta_{\mu\nu}\bar{\square}  - \bar{\nabla}_\mu\bar{\nabla}_\nu \right)\ln\left(-\frac{\bar{\square}}{\mu^2}\right)\mathcal{R}^{(1)}[h]\nonumber\\
	&\quad -\bar{\beta} \left( \delta^\alpha_\mu \delta_{\nu}^{\beta}\bar{\square} + \eta_{\mu\nu}\bar{\nabla}^\alpha\bar{\nabla}^\beta - \delta^\alpha_\mu\bar{\nabla}^\beta\bar{\nabla}_\nu - \delta^\alpha_\nu\bar{\nabla}^\beta\bar{\nabla}_\mu \right)\ln\left(-\frac{\bar{\square}}{\mu^2}\right)\mathcal{R}^{(1)}_{\alpha\beta}[h].
\end{align}

Let us now solve Eq.~\eqref{eq:linear eom} outside the dust shells, then the linear part of the metric can be obtained as
\begin{equation}
	\label{eq:linear order g}
	h_{tt}^{(1)} = h_{rr}^{(1)} = \frac{2GM}{r}.
\end{equation}
By plugging Eq.~\eqref{eq:linear order g} into Eq.~\eqref{eq:quadratic eom},
we obtain the quadratic part of the metric as
\begin{align}
	\label{}
	h^{(2)}_{tt} &= \frac{768\pi(\bar{\alpha}+\bar{\beta})G^2MN}{R^2(N+1)(2N+1)r}\sum_{n=1}^{N}\frac{R_n^2}{r^2-R_n^2},\label{eq:2nd htt}\\
	h^{(2)}_{rr} &= \frac{4G^2M^2}{r^2} + \frac{768\pi\bar{\alpha}G^2MN}{R^2(N+1)(2N+1)r}\sum_{n=1}^{N}\frac{R_n^2(R_n^2-3r^2)}{(r^2-R_n^2)^2}.\label{eq:2nd hrr}
\end{align}
Note that $h^{(2)}_{tt}$ and $h^{(2)}_{tt}$ exhibit divergences near each dust shell, rendering the perturbative approach invalid in those region.
The divergences are inevitable due to the discontinuity of the matter source.
To circumvent this issue, we restrict our solutions to the asymptotic region  where $r\gg R$, such that Eqs.~\eqref{eq:2nd htt} and \eqref{eq:2nd hrr} become
\begin{align}
    h_{tt}^{(2)} = &\frac{128\pi (\bar{\alpha}+\bar{\beta}) G^2 M}{r^3} + \frac{128\pi (\bar{\alpha}+\bar{\beta}) G^2MR^2(3N^2+3N-1)}{5N^2r^5},\label{eq:corrected_tt_RnMn}\\
    h_{rr}^{(2)} = &\frac{4G^2M^2}{r^2}- \frac{384\pi \bar{\alpha} G^2 M}{r^3}- \frac{128\pi \bar{\alpha} G^2MR^2(3N^2+3N-1)}{N^2r^5} .\label{eq:corrected_rr_RnMn}
\end{align}
Combining Eqs.~\eqref{eq:linear order g}, \eqref{eq:corrected_tt_RnMn}, and \eqref{eq:corrected_rr_RnMn}, the metric \eqref{eq:metric perturb} can be written
as
\begin{align}
	g_{tt} &= - 1 + \frac{2G M}{r} + \frac{128\pi (\bar{\alpha}+\bar{\beta}) G^2 M}{r^3} + \frac{128\pi (\bar{\alpha}+\bar{\beta}) G^2MR^2(3N^2+3N-1)}{5N^2r^5},\label{eq:gtt}\\
	g_{rr} &= 1 + \frac{2GM}{r} + \frac{4G^2M^2}{r^2}- \frac{384\pi \bar{\alpha} G^2 M}{r^3}- \frac{128\pi \bar{\alpha} G^2MR^2(3N^2+3N-1)}{N^2r^5}.\label{eq:grr}
\end{align}
The internal structure of the $N$ dust shells is prominent through the $r^{-5}$ corrections, where the number of dust shells $N$ for the fixed $R$ appears as quantum hair.
In the limit $N\to\infty$ for the solid sphere, Eqs.~\eqref{eq:gtt} and \eqref{eq:grr} reduce to
\begin{align}
	g_{tt} &= - 1 + \frac{2G M}{r} + \frac{128\pi (\bar{\alpha}+\bar{\beta}) G^2 M}{r^3} + \frac{384\pi (\bar{\alpha}+\bar{\beta}) G^2MR^2}{5r^5},\label{eq:gtt N inf}\\
	g_{rr} &= 1 + \frac{2GM}{r} + \frac{4G^2M^2}{r^2}- \frac{384\pi \bar{\alpha} G^2 M}{r^3}- \frac{128\pi \bar{\alpha} G^2MR^2}{r^5},\label{eq:grr N inf}
\end{align}
which are compatible with the quantum-corrected metric for a solid sphere up to $r^{-3}$ in Ref~\cite{Perrucci:2024qrr} and up to $r^{-5}$ in Refs.~\cite{Calmet:2019eof,Calmet:2021stu}.

\section{Quantum-corrected deflection angle}
\label{sec:deflection_angles}
To investigate quantum-corrected gravitational lensing effects, we begin with a geodesic equation for a null particle moving in the equatorial plane, \textit{i.e.}, $\theta = \frac{\pi}{2}$.
The line element for a static and spherically symmetric spacetime, restricted to this plane, is
    \begin{equation}\label{eq:metric}
      \dd s^2 = -A(r)\dd t^2 + B(r)\dd r^2 + r^2\dd\phi^2,
    \end{equation}
where $A(r) = -g_{tt}(r)$ and $B(r) = g_{rr}(r)$ defined in Eqs.~\eqref{eq:gtt} and \eqref{eq:grr}.
Now, the geodesic equation is given as $u^\nu\nabla_{\nu}u^{\mu} = 0$, where $u^\mu$ is the four-velocity of the null particle propagating on the metric~\eqref{eq:metric}.
Then, we can identify two constants of motion: the conserved energy $E$ and angular momentum $J$ as
\begin{equation}
	\label{}
	A(r)\dv{t}{\lambda}  = E,\quad r^2\dv{\phi}{\lambda} = J,
\end{equation}
where $\lambda$ is an affine parameter along the geodesic.
Imposing the null condition $u_\mu u^\mu=0$ leads to
    \begin{equation}\label{eq:r_to_phi_equation}
      \left( \dv{r}{\phi} \right)^2 = \frac{r^4}{B(r)} \left[ \frac{1}{b^2 A(r)} - \frac{1}{r^2} \right]
    \end{equation}
with $b = \frac{J}{E}$ denoting the impact parameter.

Next, we consider a light ray starting from the source, passing near the lens with the closest approach distance $r_0$, and then moving away to infinity.
At $r=r_0$, the left-hand side of Eq.~\eqref{eq:r_to_phi_equation} vanishes so that $b^2 = \frac{r_0^2}{A(r_0)}$.
The azimuthal shift of the light ray as it travels from $r_0$ to infinity is given by~\cite{Weinberg:1972kfs}
    \begin{equation}\label{eq:delta_phi}
      I(r_0) = \int_{r_0}^{\infty} \dv{\phi}{r} \dd r = \int_{r_0}^{\infty} \frac{r_0\sqrt{A(r)B(r)}}{r \sqrt{r^2A(r_0) - r_0^2 A(r)}}\dd r,
    \end{equation}
and then the deflection angle can be defined as
    \begin{equation}\label{eq:deflection_angle}
      \widehat{\alpha}(r_0) = 2 I(r_0) - \pi.
    \end{equation}
Plugging Eqs.~\eqref{eq:gtt} and \eqref{eq:grr} into Eq.~\eqref{eq:deflection_angle}, we obtain the quantum-corrected deflection angle for the lens composed of $N$ dust shells:
    \begin{align}\label{eq:deflection_angle_RnMn}
        \widehat{\alpha}_{N} = &\frac{4GM}{r_0} + \frac{4G^2 M^2}{r_0^2} \left( \frac{15}{16}\pi - 1 \right) + \frac{256\pi \bar{\beta} G^2M}{r_0^3} + \frac{1024\pi \bar{\beta} G^2MR^2(3N^2 + 3N -1)}{15N^2 r_0^5}.
    \end{align}
Here, we retain terms only up to second order in the gravitational constant $G$, compatible with the metric perturbation~\eqref{eq:metric perturb}.
In Eq.~\eqref{eq:deflection_angle_RnMn}, the first two terms represent the classical contribution arising from the Schwarzschild geometry and the last two terms come from quantum corrections.
In particular, the last term explicitly demonstrates the dependence of the deflection angle on the number of dust shells.
Additionally, the deflection angle in the continuum limit of $N$ can also be obtained by taking the limit $N\to \infty$, which yields
    \begin{align}\label{eq:deflection_angle_sp}
        \widehat{\alpha}_{\infty} = \frac{4GM}{r_0} + \frac{4G^2 M^2}{r_0^2} \left( \frac{15}{16}\pi - 1 \right) + \frac{256\pi \bar{\beta} G^2M}{r_0^3} + \frac{1024\pi \bar{\beta} G^2MR^2}{5r_0^5}.
    \end{align}
Therefore, when the quantum corrections are taken into account, compact objects of the identical mass and radius but distinct internal structures can be distinguished through their quantum-corrected deflection angles.

\section{Quantum-corrected lensing observables}
\label{sec:numerics}
\begin{figure}[t]
\centering
\begin{tikzpicture}
	\draw[thick,dotted] (0,4.5) -- (0,0);
	\draw[fill=black] (5,0) circle(2pt) node[below] {$L$};
	\draw[fill=black] (15,0) circle(2pt);
	\draw[fill=black] (0,1.5) circle(2pt);
	\draw[fill=black] (0,4.5) circle(2pt);
	\draw[thick, rounded corners=20] (0,1.5) -- (5,3)  -- (15,0);
	\draw[thick,dotted] (15,0) -- (0,4.5);
	\draw[thick,dotted] (15,0) -- (0,1.5);
	\draw[thick,dotted] (0,1.5) -- (7,3.6);
	\draw (12.5,0) arc(180:163.301:2.5) node[midway, left]{$\widehat{\theta}$};
	\draw (10,0) arc(180:174.289:5) node[midway, left]{$\widehat{\beta}$};
	\draw (4,2.7) arc(196.699:163.301:{sqrt(1.09)}) node[midway, left]{$\widehat{\alpha}$};
	\draw (6,2.7) arc(-16.699:16.699:{sqrt(1.09)}) node[midway, right]{$\widehat{\alpha}$};
	\draw[dashed] (5,0) -- (4.17431,2.75229) node[midway, left]{$b$};
	\draw[dashed] (5,0) -- (5.82569,2.75229) node[midway, right]{$b$};
	\draw[thick,<->] (0,0) -- (5-0.0705556,0) node[midway, below]{$D_{ds}$};
	\draw[thick,<->] (5+0.0705556,0) -- (15-0.0705556,0) node[midway, below]{$D_{d}$};
	\draw[thick,<->] (0,-.55) -- (15-0.0705556,-.55) node[midway, below]{$D_{s}$};
	\node[above] at (15,0) {$O$};
	\node[left] at (0,1.5) {$S$};
	\node[left] at (0,4.5) {$I$};
\end{tikzpicture}
\caption{The observer $O$, lens $L$, source $S$, and image $I$ are shown in the lens diagram.
The source and image are at angles $\widehat{\beta}$ and $\widehat{\theta}$ on the image plane, respectively, and the lens is positioned between the observer and the image plane. The light ray emitted from the source is bent by the deflection angle $\widehat{\alpha}$ along with the impact parameter $b$.
The distance from the image plane to the lens is denoted by $D_{ds}$ and the distance from the image plane to the observer is $D_{s}$, where the observer and the lens are separated by a distance $D_{d}$.
}
\label{fig:schematics}
\end{figure}
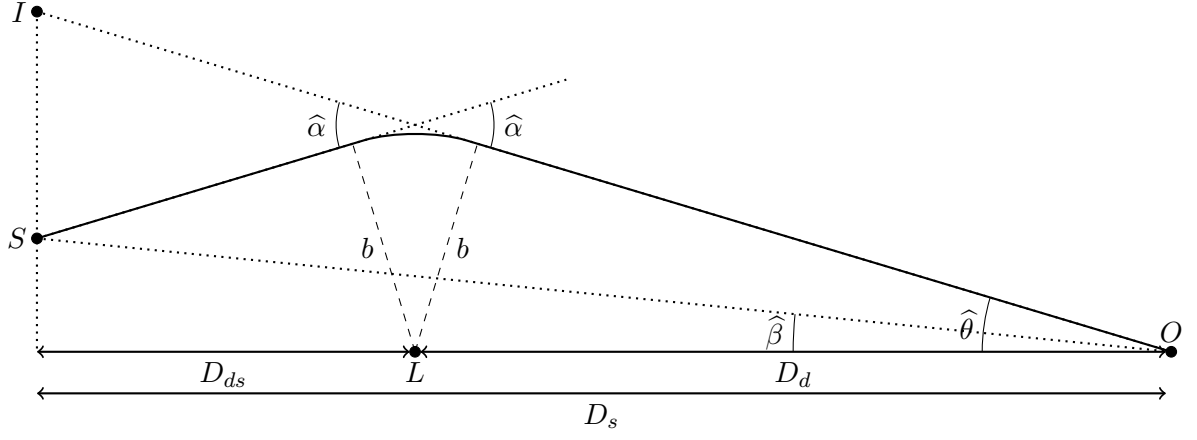
In this section, we numerically study the Einstein angle and the magnification of an image.
In Fig.~\ref{fig:schematics}, the light ray denoted by a solid curve is emitted from the source $S$ and reaches the observer $O$.
As the ray passes near the lens $L$, it is deflected by the angle~\eqref{eq:deflection_angle_RnMn}, causing the observer to perceive the image $I$ at the angle $\widehat{\theta}$ on the image plane.
Then, the lens equation can be written as~\cite{Virbhadra:1999nm}
    \begin{equation}\label{eq:lens_equation}
        \tan\widehat{\beta} = \tan\widehat{\theta} - \frac{D_{ds}}{D_{s}} \left[ \tan\widehat{\theta} + \tan (\widehat{\alpha}-\widehat{\theta}) \right].
    \end{equation}
The angle $\widehat{\theta}$ in Eq.~\eqref{eq:lens_equation} can be expressed in terms of $r_0$ as
    \begin{equation}\label{eq:impact_parameter}
	   \sin \widehat{\theta} = \frac{b}{D_d} = \frac{r_0}{D_d\sqrt{A(r_0)}},
    \end{equation}
which allows the lens equation~\eqref{eq:lens_equation} to be rewritten entirely in terms of $r_0$.
For a given source position, the light can also be deflected below the lens, producing another image at $\widehat{\theta}<0$.
In our analysis, however, we neglect this image and focus solely on the image located at $\widehat{\theta}>0$ for simplicity.

When a bundle of rays is emitted from the source, each ray is deflected depending on the distance of its closest approach $r_0$, causing the cross section of the bundle to become distorted.
\begin{figure}[t]
	\centering
	\subfigure[~The tangential magnification $\mu_{\rm t}$]{\includegraphics[width=0.475\textwidth]{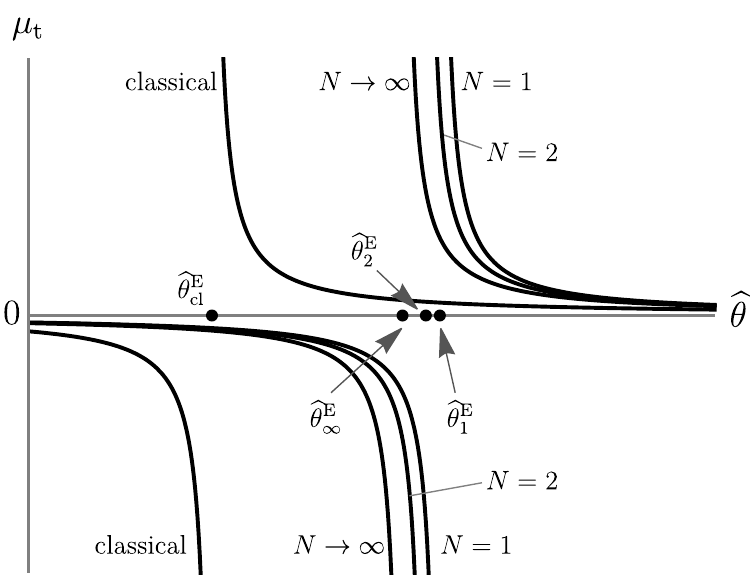}\label{fig:mu_t}}\qquad
	\subfigure[~The radial magnification $\mu_{\rm r}$]{\includegraphics[width=0.475\textwidth]{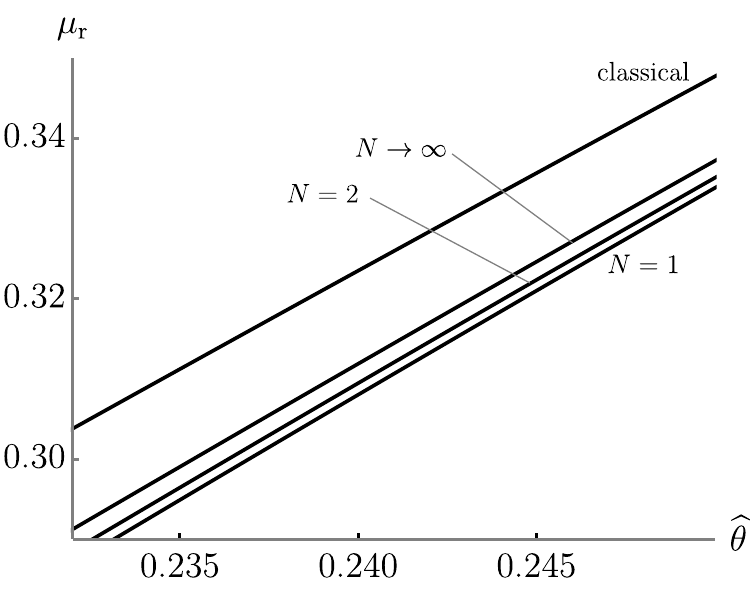}\label{fig:mu_r}}
	\caption{The tangential and radial magnifications are shown in Figs.~\ref{fig:mu_t} and \ref{fig:mu_r}, respectively.
        In Fig.~\ref{fig:mu_t}, the classical value of Einstein angle for any $N$ is $\widehat{\theta}^{\rm E}_{\rm cl} \approx 0.296071$ for $r_0 \approx 1.347775$.
		Quantum mechanically, when $N=1$, $\widehat{\theta}^{\rm E}_1 \approx 0.297399$ for $r_0 \approx 1.352829$.
		If $N=2$, $\widehat{\theta}^{\rm E}_2 \approx 0.297318$ for $r_0 \approx 1.352507$. Extremely, if $N\to\infty$, $\widehat{\theta}^{\rm E}_\infty \approx 0.297182$ for $r_0 \approx 1.351969$.
		In these calculations, we set $G=1$, $R=1$, $M=0.1$, $D_{d}=5$, and $D_{ ds}=35$ in which $D_{ s} = D_{ d} + D_{ ds}$.
	}\label{fig:mu}
\end{figure}
For a spherically symmetric lens, the magnification of the image is given by~\cite{Virbhadra:1999nm}
    \begin{equation}\label{eq:mu_tot}
        \mu_{\rm tot} = \left( \frac{\sin\widehat{\beta}}{\sin\widehat{\theta}} \dv{\widehat{\beta}}{\widehat{\theta}} \right)^{-1},
    \end{equation}
in which the tangential and radial magnifications are
    \begin{equation}\label{eq:mu_tr}
        \mu_{\rm t} = \left( \frac{\sin\widehat{\beta}}{\sin\widehat{\theta}} \right)^{-1},\quad \mu_{\rm r} = \left( \dv{\widehat{\beta}}{\widehat{\theta}} \right)^{-1}.
    \end{equation}
The source, lens, and observer are aligned for $\widehat{\beta}=0$, leading to the divergence of $\mu_{\rm tot}$ and $\mu_{\rm t}$ in Eqs.~\eqref{eq:mu_tot} and \eqref{eq:mu_tr}.
The corresponding $\widehat{\theta}$ for $\widehat{\beta}=0$ is known as the Einstein angle $\widehat{\theta}^{\rm E}$.
In Figs.~\ref{fig:mu_t} and \ref{fig:mu_r}, the tangential and radial magnifications $\mu_{\rm t}$ and $\mu_{\rm r}$ are plotted as functions of $\widehat{\theta}$, where the classical results are obtained by setting $\bar{\alpha}=\bar{\beta} = 0$.

In Fig.~\ref{fig:mu_t}, the quantum-corrected Einstein angle shifts away from its classical value.
The quantum-corrected Einstein angle is shown to be larger than the classical Einstein angle for all $N$.
The largest Einstein angle occurs for $N=1$ and the smallest one is attained for $N\to\infty$.
On the other hand, the tangential magnification $\mu_{\rm t}$ normally diverges at $\widehat{\theta}=\widehat{\theta}^{\rm E}$.
If  $\widehat{\theta}>\widehat{\theta}^{\rm E}$, it is positive while it becomes negative for $\widehat{\theta}<\widehat{\theta}^{\rm E}$.
The positive region of $\mu_{\rm t}$ indicates the primary image which is brighter than the source, whereas the negative region corresponds to the secondary image which is dimmer than the source.
For a given $\widehat{\theta}$, the quantum-corrected $\mu_{\rm t}$ decreases as $N$ increases, although it is always greater than the classical $\mu_{\rm t}$ which is the minimum.
Additionally, the radial magnification $\mu_{\rm r}$ in Fig.~\ref{fig:mu_r} increases monotonically with $\widehat{\theta}$.
For any $N$, the classical $\mu_{\rm r}$ is always larger than the quantum-corrected $\mu_{\rm r}$.

\section{conclusion and discussion}
\label{sec:conclusion}
In order to investigate the internal structure of the star composed of $N$ dust
shells, we obtained the quantum-corrected metric using the effective field theory approach.
The quantum-corrected metric exhibits $r^{-5}$ corrections, which depend on the number of dust shells $N$ for the fixed $R$, reflecting the presence of quantum hair.
Our study shows that quantum effects modify the deflection angle and Einstein angle in gravitational lensing.
The deviation between the classical and quantum-corrected results is maximized for $N=1$ and decreases as $N$ increases, approaching a certain limit as $N\to \infty$; however,
the quantum-corrected deflection angle and Einstein angle are found to be
eventually larger than the classical ones.
In conclusion, the internal structures of compact objects can be distinguished by quantum hair through the quantum-corrected lensing observables.
Although such quantum corrections to the lensing observables are extremely small and may be practically undetectable, it is worth emphasizing that these results can be derived from first principles in a model-independent manner.

Our results may raise some questions.
Firstly, how do the quantum-corrected deflection angle and Einstein angle become always larger than the classical ones?
The essential reason lies in the effective mass in Eq.~\eqref{eq:gtt}.
The effective mass can be defined as the classical mass $M$ supplemented by quantum-corrected terms such as $M_{\rm eff}(r)= M \left( 1+\frac{64\pi (\bar{\alpha}+\bar{\beta})G}{r^2}+\frac{64\pi (\bar{\alpha}+\bar{\beta}) GR^2(3N^2+3N-1)}{5N^2r^4} \right)$.
Since these quantum contributions are positive definite, the effective mass is always heavier than $M$.
As a result, in the quantum-corrected lensing effect, the light ray interacts with the heavier effective mass of dust shells, leading to a stronger deflection of light compared to the classical case.
Secondly, how do the quantum-corrected deflection angle and Einstein angle decrease with increasing $N$?
As $N$ increases, the quantum-corrected terms in the effective mass $M_{\rm eff}(r)$ gradually diminish.
Consequently, the effective mass of dust shells decreases with increasing $N$, leading to a weaker interaction between the light ray and the dust shells, which results in a smaller deflection angle and Einstein angle for a large $N$.

A final comment is in order. In the previous studies~\cite{Calmet:2017qqa,Calmet:2019eof,Calmet:2020tlj,Xiao:2021zly,Cheong:2023oik,Perrucci:2024qrr,Calmet:2023met},
the quantum-corrected metric has been obtained around the Schwarzschild background
instead of the flat metric.
This is a crucial distinction from our calculations which employ perturbations around the flat metric.
In the Barvinsky-Vilkovisky method, the effective action should be derived through the curvature expansion around flat spacetime.
Based on the spirit of the Barvinsky-Vilkovisky method~\cite{Barvinsky:1983vpp,Barvinsky:1985an,Barvinsky:1987uw,Barvinsky:1990up,Barvinsky:1990uq}, we obtained the quantum-corrected metric by taking the background as the flat metric.
Nevertheless, up to quadratic order in $G$, the resulting quantum-corrected metrics~\eqref{eq:gtt} and \eqref{eq:grr} happen to be coincident regardless of whether the Schwarzschild metric or flat metric is chosen as the background.
This means that, at quadratic order in $G$, perturbations around the Schwarzschild background are fortunately equivalent to those around flat spacetime.
However, when considering terms beyond quadratic order in $G$, the perturbative expansion may differ depending on the choice of background.

\acknowledgments
We would like to thank Sunghyun Kang and Stefano Scopel for exciting discussions.
This research was supported by Basic Science Research Program through the National Research Foundation of Korea(NRF) funded by the Ministry of Education through the Center for Quantum Spacetime (CQUeST) of Sogang University (No. RS-2020-NR049598).
This work was supported by the National Research Foundation of Korea(NRF) grant funded by the Korea government(MSIT).(No. RS-2022-NR069013)


\bibliographystyle{JHEP}       

\bibliography{references}

\end{document}